\pdfoutput=1
\documentclass[11pt,twoside,a4paper,cmspaper,final,collab]{cms-tdr}
\def\svnVersion{3e9cab1}\def\svnDate{2023/12/28}\def\cmsCernNoTag{CERN-EP-2023-163}\def\cmsCernDate{\today}\def\cmsMessage{Published in the European Physical Journal C as \href{http://dx.doi.org/10.1140/epjc/s10052-023-12268-2}{\doi{10.1140/epjc/s10052-023-12268-2}.}}
\begin{document}\cmsNoteHeader{LUM-21-001}

\newlength\cmsTabSkip\setlength{\cmsTabSkip}{1ex}
\ifthenelse{\boolean{cms@external}}{%
    \providecommand{\cmsleft}{upper\xspace}
    \providecommand{\cmsright}{lower\xspace}
    \providecommand{\cmsLeft}{Upper\xspace}
    \providecommand{\cmsRight}{Lower\xspace}
}{%
    \providecommand{\cmsleft}{left\xspace}
    \providecommand{\cmsright}{right\xspace}
    \providecommand{\cmsLeft}{Left\xspace}
    \providecommand{\cmsRight}{Right\xspace}
}

\newcommand{\pp}{\ensuremath{\Pp\Pp}\xspace}
\newcommand{\ZTOMM}{\ensuremath{\PZ\to\PGm\PGm}\xspace}
\newcommand{\ZTOMPMM}{\ensuremath{\PZ\to\MM}\xspace}
\newcommand{\mll}{\ensuremath{m_{\PGm\PGm}}\xspace}
\newcommand{\sZ}{\ensuremath{\sigma^\PZ}\xspace}
\newcommand{\sZfid}{\ensuremath{\sigma^\PZ_\text{fid}}\xspace}
\newcommand{\effXY}{\ensuremath{\epsilon^{\PGm}_{x|y}}\xspace}
\newcommand{\effZID}{\ensuremath{\epsilon^{\PZ}_\mathrm{ID}}\xspace}
\newcommand{\effMID}{\ensuremath{\epsilon^{\PGm}_\mathrm{ID}}\xspace}
\newcommand{\effMHLT}{\ensuremath{\epsilon^{\PGm}_\mathrm{HLT}}\xspace}
\newcommand{\effMIDGlo}{\ensuremath{\epsilon^{\PGm}_\mathrm{ID|Glo}}\xspace}
\newcommand{\effMGloSta}{\ensuremath{\epsilon^{\PGm}_\mathrm{Glo|Sta}}\xspace}
\newcommand{\effMStaTrk}{\ensuremath{\epsilon^{\PGm}_\mathrm{Sta|Trk}}\xspace}
\newcommand{\Run}[1]{\mbox{Run~#1}\xspace}
\newcommand{\BR}{\ensuremath{\mathcal{B}}\xspace}
\newcommand{\lowPU}{\ensuremath{\mathrm{lowPU}}\xspace}
\newcommand{\highPU}{\ensuremath{\mathrm{highPU}}\xspace}
\newcommand{\Nbkg}[1]{\ensuremath{N_{#1}^{\text{bkg}}}\xspace}
\newcommand{\Nbkgi}{\ensuremath{N_i^{\text{bkg}}}\xspace}
\newcommand{\Nsig}[1]{\ensuremath{N_{#1}^{\text{sig}}}\xspace}
\newcommand{\Nsigi}{\ensuremath{N_i^{\text{sig}}}\xspace}
\newcommand{\N}[1]{\ensuremath{N_{#1}}\xspace}
\newcommand{\Ni}{\ensuremath{N_i}\xspace}
\newcommand{\NZ}{\ensuremath{N^{\PZ}}\xspace}
\newcommand{\CHLT}{\ensuremath{C_{\mathrm{HLT}}}\xspace}
\newcommand{\CIO}{\ensuremath{c_{\mathrm{T\&P}}}\xspace}
\newcommand{\CID}{\ensuremath{C_{\mathrm{ID}}}\xspace}
\newcommand{\NPV}{\ensuremath{N_{\mathrm{PV}}}\xspace}
\newcommand{\abseta}{\ensuremath{\abs{\eta}}\xspace}
\newcommand{\npass}{\ensuremath{n^\mathrm{p}}\xspace}
\newcommand{\nfail}{\ensuremath{n^\mathrm{f}}\xspace}
\newcommand{\nbsinv}{\mbox{\ensuremath{\,\text{nb}^{-1}\,\text{s}^{-1}}}\xspace}

\cmsNoteHeader{LUM-20-001}
\title{Luminosity determination using \texorpdfstring{\PZ}{Z} boson production at the CMS experiment}

\date{\today}

\abstract{The measurement of \PZ boson production is presented as a method to determine the integrated luminosity of CMS data sets. The analysis uses proton-proton collision data, recorded by the CMS experiment at the CERN LHC in 2017 at a center-of-mass energy of 13\TeV. Events with \PZ bosons decaying into a pair of muons are selected. The total number of \PZ bosons produced in a fiducial volume is determined, together with the identification efficiencies and correlations from the same data set, in small intervals of 20\pbinv of integrated luminosity, thus facilitating the efficiency and rate measurement as a function of time and instantaneous luminosity. Using the ratio of the efficiency-corrected numbers of \PZ bosons, the precisely measured integrated luminosity of one data set is used to determine the luminosity of another. For the first time, a full quantitative uncertainty analysis of the use of \PZ bosons for the integrated luminosity measurement is performed. The uncertainty in the extrapolation between two data sets, recorded in 2017 at low and high instantaneous luminosity, is less than 0.5\%. We show that the \PZ boson rate measurement constitutes a precise method, complementary to traditional methods, with the potential to improve the measurement of the integrated luminosity.}

\hypersetup{%
pdfauthor={CMS Collaboration},%
pdftitle={Integrated luminosity determination using Z boson production at the CMS experiment},%
pdfsubject={CMS},%
pdfkeywords={CMS, luminosity, Z counting}}

\maketitle

\section{Introduction}

In the CERN LHC, during the \Run2 data-taking period in 2015--2018, about 300 million events with \PZ bosons decaying into pairs of muons were recorded by the CMS experiment.
Precision cross section measurements were performed~\cite{CMS:EWK-10-005, CMS:SMP-17-001, CMS:SMP-17-010, CMS:2022ubq, ATLAS:2016fij, LHCb:2022kph}
that provide (i) important tests of theoretical calculations~\cite{Li:2012wna, Monni:2019whf, Duhr:2021vwj}; (ii)
input to fits of the parton distribution functions (PDFs) of the proton~\cite{Alekhin:2017kpj, EXT:NNPDF-2017, Dulat:2015mca, Harland-Lang:2014zoa}; and (iii) constraints on backgrounds to searches for new physics~\cite{CMS:HIG-17-023}.

Events with a \PZ boson decaying into a pair of muons have a remarkably clean experimental signature and a large cross section that facilitates high-precision measurements. Samples of \PZ bosons are also used as standard tools for detector calibrations and efficiency studies. The precisely known \PZ boson mass and width~\cite{EXT:PDG-2022} are used to calibrate energy scales and momenta and to determine the detector resolution~\cite{CMS:EGM-17-001, CMS:MUO-16-001}.
Efficiencies for lepton triggering, reconstruction, and identification are determined using the ``tag-and-probe'' method~\cite{CMS:EWK-10-005, ATLAS:2016lqx, CMS:EGM-17-001, CMS:MUO-16-001}.

The large Drell--Yan (DY) cross section for the production of \PZ bosons, and the possibility of simultaneously determining both the yield and the detection efficiency in situ, \ie, from the same event sample, make the process useful for precision measurements of the integrated luminosity. This was discussed before the start of the LHC~\cite{Dittmar:1997md}.
During LHC operation, measurements of the \PZ boson rate already proved to be a useful and independent method for the LHC machine operators and experiments to monitor the relative instantaneous luminosity delivered to the ATLAS and CMS experiments~\cite{Dubourg:2019mkt}.
The use of \PZ boson production as a measure of relative luminosities was also explored by the ATLAS experiment~\cite{ATLAS:2022hro}.

Both muons from the \PZ boson decay are detectable within the fiducial volume of the CMS detector in about one third of the \PZ boson events. The fiducial \PZ boson cross section in proton-proton (\pp) collisions at 13\TeV has been measured to be~\cite{CMS:SMP-17-010}
\begin{linenomath}\begin{equation}
    \sZ \BR(\ZTOMPMM) = 694 \pm 6 \syst \pm 17 \lum\unit{pb}.
\end{equation}\end{linenomath}
Theoretical predictions are available up to next-to-next-to-next-to-leading order (N\textsuperscript{3}LO)~\cite{Duhr:2021vwj} in quantum chromodynamics (QCD). Electroweak corrections, including mixed QCD-electro\-weak corrections, are also available~\cite{Li:2012wna,Bonciani:2021zzf,Delto:2019ewv}.
The current uncertainty in the prediction of the fiducial cross section is about 3\%, and mainly originates from limited knowledge of proton PDFs and higher-order corrections~\cite{Monni:2019whf}.
Within this uncertainty, the integrated luminosity can be directly determined from the measured number of \PZ bosons corrected for efficiencies.

{\tolerance=800
In practice, precision luminosity calibrations at the LHC are obtained from van der Meer (vdM) scan data~\cite{vanderMeer:296752, Grafstrom:2015foa, Aaij:2014ida, Abelev:2014epa, ATLAS:2016fhk, CMS:LUM-17-003, ATLAS:2022hro}, which are more precise than the theory predictions for the Z boson cross section.
In vdM scans, which are performed at low instantaneous luminosity with zero crossing angle between the two beams, the two beams are separated in two orthogonal directions transverse to the parallel beam axes.
In each scan step, for a given beam separation, the event rate measured in the luminosity detectors is recorded to determine the beam overlap area. Together with the beam currents and the measured head-on collision rate, a luminosity calibration constant, referred to as the visible cross section, is determined.
A full vdM scan campaign takes about six hours per experiment and is usually performed once per year, with specifically configured beams to maximize the accuracy and precision of the measurement. A detailed description of vdM scans is reported in Ref.~\cite{CMS:LUM-17-003}.
\par}

{\tolerance=800
The most precise integrated luminosity measurement in CMS to date, achieved for the 2016 data-taking period,
has a total uncertainty of 1.2\%~\cite{CMS:LUM-17-003}.
Roughly half of the total uncertainty is due to the luminosity integration over the full year of data taking.
This uncertainty, in turn, is composed of the uncertainty in the extrapolation of the visible cross section obtained in the vdM scan to standard data-taking conditions at high instantaneous luminosity, and the uncertainty in the integration of the instantaneous luminosity over time, obtained from comparisons between different luminosity measurements. In the 2017 data, presented in this paper, the average number of collisions per bunch crossing, usually referred to as pileup, was 32~\cite{CMS:JME-18-001}. In \Run2, peak instantaneous luminosities as high as 20\nbsinv were reached, corresponding to a pileup of more than 50.
For the high-luminosity LHC (HL-LHC), a pileup of up to 200 is expected~\cite{Dainese:2703572} likely leading to an increase in uncertainty with the conventional methods due to the larger extrapolation.
\par}

In this paper, we explore an approach originally proposed in Ref.~\cite{Salfeld-Nebgen:2018bdp}. The measurement of the \PZ boson rate is used as an alternative method for the extrapolation and integration of the luminosity calibration. The \PZ boson counting complements conventional luminosity measurements obtained from the CMS luminosity systems, which are taken as reference luminosity.
The fiducial \PZ boson production cross section is defined as $\sZfid=\NZ/\lumi$, where \NZ stands for the efficiency-corrected number of reconstructed \PZ boson events and \lumi for the integrated luminosity.
Since \sZfid is identical for all data sets of the same center-of-mass energy, the ratio of \NZ for two data sets can be used to transfer the luminosity calibration from one data set to another, without input from theoretical predictions or precise knowledge of \sZfid.
For the first time, a full quantitative uncertainty analysis of the use of \PZ bosons for the integrated luminosity measurement is performed.

We choose two independent data sets of \PZ boson events, both recorded by the CMS experiment in 2017:\ a data set with a bunch luminosity corresponding to about three \pp collisions per bunch crossing, referred to in the following as ``\lowPU''; and the bulk of CMS \pp collision data recorded in 2017 with a typical pileup of 32, denoted as ``\highPU''.
The luminosity of the \lowPU data is used to determine that of the \highPU data via the relation
\begin{linenomath}\begin{equation}
\label{eqn:lumiratio}
    \lumi_\highPU =  \frac{\NZ_\highPU}{\NZ_\lowPU} \lumi_\lowPU.
\end{equation}\end{linenomath}
For both sets of data, the individual trigger and selection efficiencies are determined in situ, in intervals of 20\pbinv, thus enhancing the sensitivity to possible variations due to changes in beam conditions or detector response as a function of time.
Using the integrated luminosity for the \lowPU data, which has an uncertainty of 1.7\%~\cite{CMS:LUM-17-004}, the integrated luminosity $\lumi_\highPU$ and its uncertainty are determined from Eq.~\eqref{eqn:lumiratio}, and compared with the result from the conventional integrated luminosity measurement. Due to the cleaner signature, better resolution, and smaller backgrounds, the analysis of \PZ boson decays into muons is more accurate than electrons.
In this paper, only the decays of \PZ bosons into muons are used.

The paper is structured as follows.
After a brief outline of the CMS detector in Section~\ref{sec:cms}, the analysis of the \PZ boson event sample is described in Section~\ref{sec:evtsel}.
The reconstructed number of \PZ bosons and the trigger and selection efficiencies are extracted from fits to the data.
Acceptance corrections and correlations between the efficiencies for the muon track components and among the two muon tracks are studied as a function of pileup.
Subsequently, in Section~\ref{sec:results}, the luminosity information obtained from \PZ boson counting is compared with the results from conventional luminosity measurements.
In Section~\ref{sec:outlook}, the benefits and advantages of \PZ boson counting for luminosity measurements are discussed.
The paper concludes with a summary in Section~\ref{sec:summary}.

\section{The CMS detector}
\label{sec:cms}

The central feature of the CMS apparatus is a superconducting solenoid of 6\unit{m} internal diameter, providing a magnetic field of 3.8\unit{T}.
Within the magnet volume are a silicon pixel and strip tracker, a lead tungstate crystal electromagnetic calorimeter (ECAL), and a brass and scintillator hadron calorimeter, each composed of a barrel and two endcap sections.
Forward calorimeters extend the pseudorapidity ($\eta$) coverage provided by the barrel and endcap detectors.
The muon system consists of gas-ionization detectors embedded in the steel flux-return yoke outside the solenoid.
A more detailed description of the CMS detector, together with a definition of the coordinate system used and the relevant kinematic variables, is reported in Ref.~\cite{CMS:Detector-2008}.

The silicon tracker measures charged particles in the pseudorapidity range $\abseta<3.0$~\cite{CMS:TRK-11-001, CMS:Pixel-2021}.
An iterative approach is used to build tracker tracks, executing a sequence of tracking algorithms, each with slightly distinct logic~\cite{CMS:MUO-16-001}.
Muons are measured in the range \mbox{$\abseta<2.4$}, with detection planes made using three technologies:\ drift tubes, cathode strip chambers, and resistive plate chambers.
Matching muons to tracks measured in the silicon tracker results in a relative transverse momentum (\pt) resolution of 1\% in the barrel and 3\% in the endcaps, for muons with \pt of about 100\GeV~\cite{CMS:MUO-16-001}.
The particle-flow (PF) algorithm~\cite{CMS:PRF-14-001} reconstructs and identifies each individual particle in an event, combining information from the various CMS detector components.
Jets are clustered using the anti-\kt jet finding algorithm~\cite{EXT:AK-2008, EXT:FastJet-2012} with the tracks assigned to candidate vertices as inputs, and the associated missing transverse momentum \ptmiss, taken as the negative vector \pt sum of those jets~\cite{CMS:JME-17-001}.
The primary vertex (PV) is taken to be the vertex with the largest $\sum\pt^2$ of its associated tracks, as described in Section 9.4 of Ref.~\cite{CMS:TDR-15-02}.

{\tolerance=800
Events of interest are selected using a two-tiered trigger system.
The first level (L1), comprised of custom hardware processors, uses information from the calorimeters and muon detectors to select events at a rate of around 100\unit{kHz} within a fixed latency of 4\mus~\cite{CMS:TRG-17-001}.
The second level, known as the high-level trigger (HLT), consists of a farm of processors running a version of the full event reconstruction software optimized for fast processing, and reduces the event rate to around 1\unit{kHz} before data storage~\cite{CMS:TRG-12-001}.
\par}

During LHC \Run2, the main CMS luminosity subdetectors (luminometers) were the silicon pixel detector, the hadron forward calorimeter (HF), the pixel luminosity telescope (PLT)~\cite{CMSBRIL:2022qin}, and the fast beam conditions monitor (BCM1F)~\cite{Guthoff:2019cil}.
A separate data acquisition system is used to collect and store HF, PLT, and BCM1F data, as well as LHC beam-related data.
A more detailed description of the CMS luminosity system is reported in Ref.~\cite{CMS:LUM-17-003}.
For all comparisons in this paper, the reference integrated luminosity is obtained with the CMS luminometers, calibrated as described in Ref.~\cite{CMS:LUM-17-004} and using offline-calibrated corrections for the afterglow effects in the HF luminosity measurement.

{\tolerance=1400
The analysis described in this paper is largely independent of Monte Carlo (MC) simulations.
However, MC simulations are used for two purposes:\ to determine the expected DY invariant mass distribution of the signal measured in the CMS detector;
and to study possible biases in the pileup-dependent measurement of the muon track-finding efficiencies.
Simulated event samples of the DY process, \mbox{$\PZ/\PGg^\ast \to \Pell\Pell$}, are produced at leading order using the \MGvATNLO (v2.6.5)~\cite{EXT:MG5aMCatNLO-2014} generator, interfaced with \PYTHIA (v8.240)~\cite{EXT:Pythia-2015} for the parton shower simulation.
The parameters describing the modeling of the parton shower and underlying event are based on the CP5 tune~\cite{CMS:GEN-17-001}.
The generated MC events are passed through a full simulation of the detector using \GEANTfour~\cite{EXT:GEANT4-2002}.
\par}

\section{The \texorpdfstring{\PZ}{Z} boson candidate selection and efficiency determination}
\label{sec:evtsel}

The events were recorded using a single-muon trigger (HLT muon) that requires at least one muon candidate with $\pt>24\GeV$ and loose isolation criteria~\cite{CMS:MUO-19-001}.
The \lowPU data were recorded using different, looser trigger configurations than those used for the \highPU data.
To obtain identical trigger configurations for the two data sets, the trigger decision in the \lowPU was recalculated from raw data using the trigger configuration of the \highPU data.

{\tolerance=800
Based on the offline reconstruction, selected muon candidates consist of an ``outer'' standalone track in the muon system, matched to an ``inner'' track reconstructed in the silicon track\-er~\cite{CMS:TRK-11-001}.
The outer track is required to have signals in at least two muon detector planes.
The inner track must have at least one valid hit in the silicon pixel detector and hits in more than five strip tracker layers.
The matching is done by comparing parameters of the two tracks propagated onto a common surface.
A combined Kalman filter fit~\cite{Fruhwirth:1987fm} is performed in which the information from the inner and outer tracks is used to obtain a ``global'' muon track.
For global muons, the inner and outer tracks are required to have \mbox{$\pt>20\GeV$}, lie within $\abseta<2.4$, and to be matched within \mbox{$\DR = \sqrt{\smash[b]{(\Delta\eta)^2 + (\Delta\phi)^2}} < 0.3$}.
Quality criteria on the global muon track fit are imposed, and it is required that the muon candidate is also reconstructed with the PF algorithm~\cite{CMS:PRF-14-001}. No requirements are imposed on the impact parameters of the muon track. Isolation criteria are omitted to maintain efficiency also at high pileup.
For muons with $\pt<200\GeV$, \ie, about 99\% of identified muon candidates, the track parameters are taken from the inner track. In other cases, the track parameters are determined by combining information from the inner and outer tracks. For all muon tracks, $\pt>25\GeV$ is required to ensure that the trigger efficiency reaches a plateau.
\par}

{\tolerance=1600
A \PZ boson candidate is identified as a pair of op\-posite-charge muons with an invariant mass of \mbox{$60 < \mll < 120\GeV$}.
At least one of the two muon candidates is required to be matched with an HLT muon within $\DR < 0.1$.
To obtain the actual number of produced \PZ bosons, the number of reconstructed and selected \PZ boson candidates, the trigger efficiency, the muon-identification efficiency, as well as the background arising from nonresonant production, are determined from dedicated fits to the data, as explained in the following.
\par}

\subsection{Trigger efficiency and signal extraction}
\label{sec:effHLT}

The trigger efficiency and the number of \PZ boson candidates are determined from fits to the invariant dimuon mass distributions of mutually exclusive sets of events with exactly one (\N1) or exactly two (\N2) selected muons matched to an HLT muon.
The observables \N1 and \N2 follow the relations
\begin{linenomath}\begin{equation}
\label{eq:N1N2}
\begin{aligned}
    \N1 &= 2 \effMHLT \big(1 - \CHLT \effMHLT \big) \effZID \NZ + \Nbkg1, \\
    \N2 &= \CHLT \big(\effMHLT\big)^2 \effZID \NZ + \Nbkg2.
\end{aligned}
\end{equation}\end{linenomath}
Here, the quantity \effMHLT refers to the HLT muon trigger efficiency.
The correction factor \CHLT accounts for the correlation between the HLT efficiencies of the two muons.
A value of $\CHLT>1$ indicates a positive correlation between the two muons, \ie, an increased probability for the second muon to pass the HLT if the first muon passes it.
The determination of \CHLT is presented in Section~\ref{sec:HLTcor}.
The terms \Nbkg1 and \Nbkg2 describe the contributions from nonresonant backgrounds.
The reconstruction efficiency \effZID is separately determined from the data, as described in Section~\ref{sec:effrec}.

\begin{figure}[!t]
\centering
\includegraphics[width=0.48\textwidth]{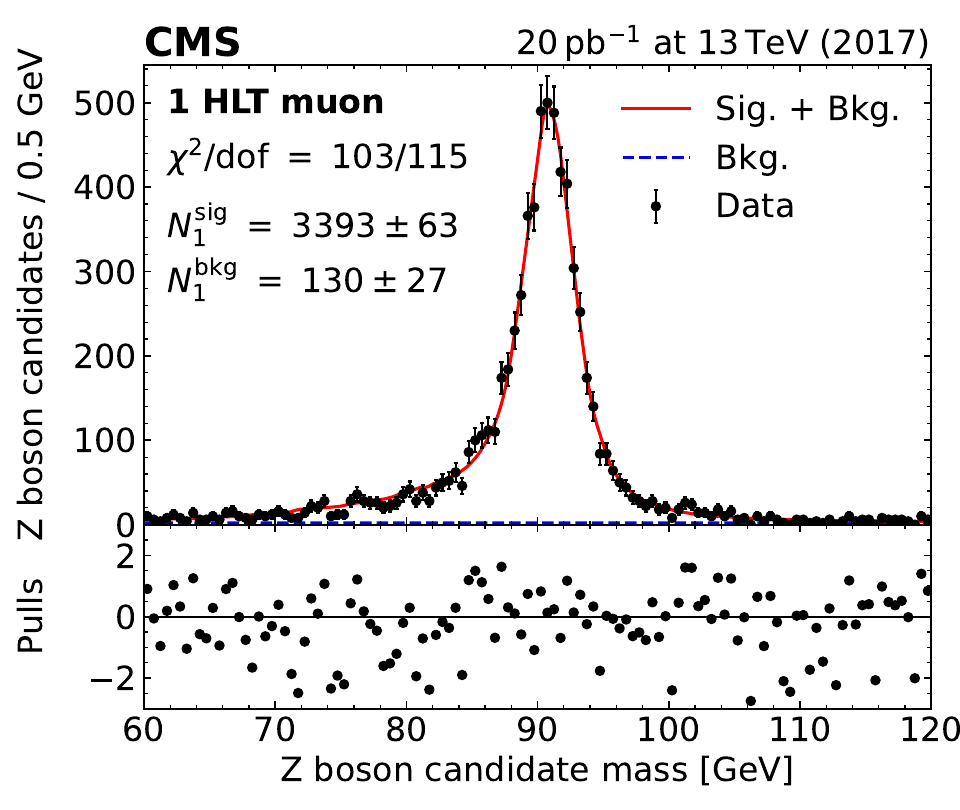}%
\hfill%
\includegraphics[width=0.48\textwidth]{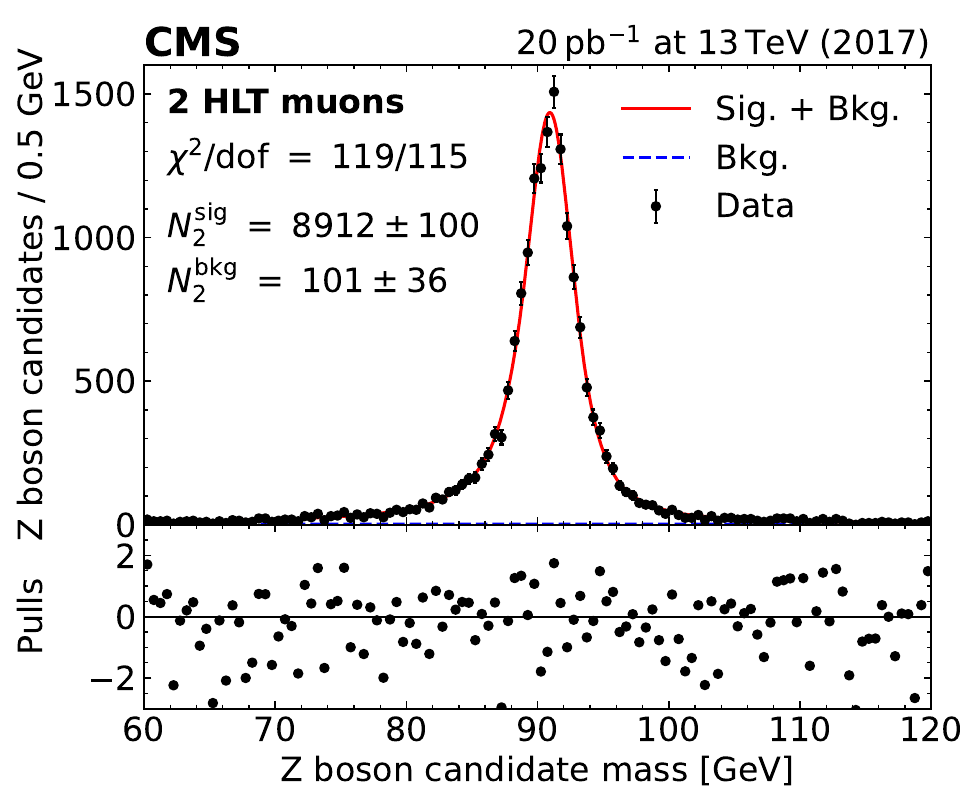}
\caption{%
    The upper panels show the reconstructed invariant mass distributions of \PZ boson candidates in a 20\pbinv sample of data for events where one (\cmsleft) or two (\cmsright) muons pass the single-muon trigger selection.
    The blue curve shows the fitted background contribution and the red curve illustrates the modeled signal-plus-background contribution.
    The error bars indicate the statistical uncertainties.
    The numbers of signal and background candidates are given by $\Nsigi=\Ni - \Nbkgi$ and \Nbkgi, respectively. Also indicated are the $\chi^2$ values per degree of freedom (dof).
    The lower panels contain the pulls of the distributions, defined as the difference between the data and the fit model in each bin, divided by the statistical uncertainty estimated from the expected number of entries given by the model.
}
\label{fig:fit}
\end{figure}

A fit is performed to two histograms binned in \mll for \PZ candidates contributing to \N1 and \N2 in which \effMHLT and \NZ are two free parameters.
In the fit, the signal is modeled by a histogram template generated from simulated \ZTOMM events, convolved with a Gaussian function to take into account muon momentum scale and resolution differences between data and simulation.
A falling exponential function is used to describe the nonresonant background.
In Fig.~\ref{fig:fit}, examples of two distributions and the results of the fits are presented.
The sample shown here corresponds to an integrated luminosity of 20\pbinv, yielding about 12\,000 \PZ boson candidates.

\subsection{Muon trigger correlation}
\label{sec:HLTcor}

The correlation between the trigger efficiencies of the two HLT muons is described by the correction factor \CHLT, as introduced in Eq.~\eqref{eq:N1N2}.
The dependence of \CHLT on the pileup is of particular interest in this analysis because it does not cancel in the ratio in Eq.~\eqref{eqn:lumiratio}, and thus constitutes an important source of systematic uncertainty.
The correlation was investigated in simulation, and it is largely understood to originate from isolation requirements in the trigger selection.

We determine \CHLT from an MC simulation sample of \ZTOMM events. As a proxy to the amount of pileup in a given event, we use the number of reconstructed PVs, \NPV, an observable that is directly accessible event-by-event in both data and simulation.
At fixed pileup, the distribution of \NPV approximately follows a Poisson distribution with a mean at about 80\% of the true pileup, as determined from DY simulation.

In the simulation, \CHLT is obtained directly, by rearranging Eq.~\eqref{eq:N1N2}, as
\begin{linenomath}\begin{equation}
    \CHLT = \frac{4 \NZ \effZID \Nsig2}{\left(\Nsig1 + 2 \Nsig2\right)^2},
\end{equation}\end{linenomath}
where \Nsig1 and \Nsig2 are the number of signal events, corresponding to $\N1-\Nbkg1$ and $\N2-\Nbkg2$ in the data.

We use data to validate the result for \CHLT obtained in the simulation. To this end, events are analyzed that are triggered independently of the muon trigger, namely by using the trigger condition $\ptmiss>120\GeV$ in which the contribution from muons is not included.
This \ptmiss trigger also records \PZ boson candidates for which the number of HLT muons is zero, and, thus, an additional relation for the number of reconstructed \PZ boson candidates with no HLT muons, denoted as \N0, is obtained,%
\begin{linenomath}\begin{equation}
    \N0 = \big(1-2\effMHLT + \CHLT (\effMHLT)^2 \big) \effZID \NZ + \Nbkg0.
\end{equation}\end{linenomath}
Together with Eq.~\eqref{eq:N1N2}, we obtain three equations for \N0, \N1, and \N2 with three unknowns, \effMHLT, \CHLT, and $\effZID\NZ$.
The correction factor \CHLT can thus be determined from the number of signal events in the three categories, each obtained from a fit.
The fits are performed separately in six bins of \NPV where the number of bins and their boundaries are chosen such that the number of events per bin are similar.

\begin{figure}[!t]
\centering
\includegraphics[width=0.48\textwidth]{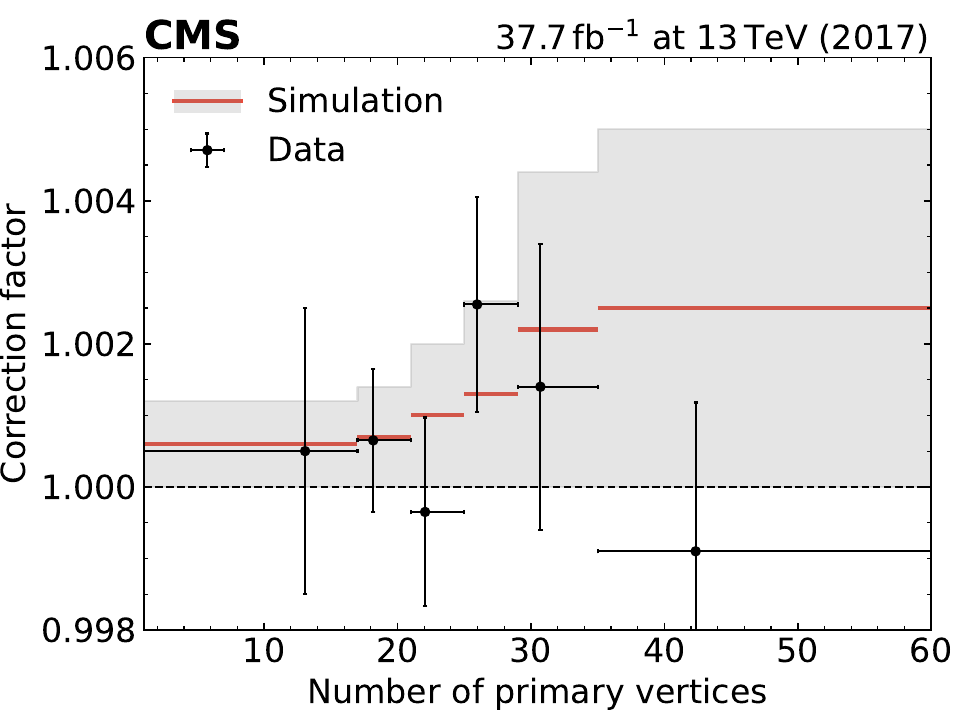}
\caption{%
    Correction factor \CHLT for the correlation between the measured muon trigger efficiencies of the two muons as a function of the number of reconstructed primary vertices, \NPV, in the simulation (lines) and the data (points).
    The data points are drawn at the mean value of \NPV in each bin of the measurement.
    The horizontal error bars on the points show the bin width, and the vertical error bars show the statistical uncertainty.
    The gray band indicates the $\pm100\%$ uncertainty in the correction factor.
}
\label{fig:corrFac_hlt_data}
\end{figure}

The result is presented in Fig.~\ref{fig:corrFac_hlt_data}.
The red lines indicate the expectation from the simulation in which \CHLT is at the level of 0.1--0.2\% above unity for \mbox{$\NPV \sim 30$}.
Within the limited statistical precision of the data, good agreement of the simulation with the data is observed.
We assign a systematic uncertainty of 100\% of the correction, which is represented by the gray band in the figure.

\subsection{Muon identification and reconstruction efficiency}
\label{sec:effrec}

{\tolerance=800
The efficiency to reconstruct a \PZ boson, \effZID, depends on the muon identification and reconstruction efficiency \effMID for each of the two muons.
In the simulation, the pileup-dependent correlation between the two identified muons is of the order of 0.01\%,
and thus \mbox{$\effZID = \CID \big( \effMID \big)^2$}. The value for $\CID \approx 1.0001$ is taken from simulation and applied as a function of \NPV.
The muon efficiency \effMID is defined independently of the HLT muon efficiency,
such that the total number of produced \PZ bosons is obtained from Eq.~\eqref{eq:N1N2}.
\par}

To determine \effMID, the following factorization ansatz is used:
\begin{linenomath}\begin{equation}
\label{eq:efficiencies}
    \effMID = \effMIDGlo \, \effMGloSta \, \effMStaTrk \, \frac{1}{\CIO},
\end{equation}\end{linenomath}
where the efficiency \effMIDGlo is the fraction of global muons that fulfill the full set of muon identification requirements;
the efficiency \effMGloSta is the global muon efficiency, given by the fraction of standalone muons that also qualify as global muon;
and the efficiency \effMStaTrk is the standalone muon efficiency, defined as the fraction of muons with good inner tracks that are matched within $\DR<0.3$ to outer standalone muon tracks with $\pt>20\GeV$ and $\abseta<2.4$.
To obtain an unbiased set of inner tracks for the measurement of the efficiency \effMStaTrk, inner tracks that are seeded from the extrapolation of outer standalone muon tracks are excluded.
The term \CIO accounts for the correlations between the efficiency terms in Eq.~\eqref{eq:efficiencies}.
The pileup dependence of the correction from \CIO between the \lowPU and the \highPU data sets is estimated from simulation to be about 0.01\%.

The efficiencies are determined from the data using a ``tag-and-probe" methodology~\cite{CMS:EWK-10-005}.
Identified muon candidates that are matched to the HLT muon are selected as ``tag''.
For each tag, a probe muon candidate of opposite charge is selected under the condition that the muon candidate pair has an invariant mass between 60 and 120\GeV.
The efficiency \effXY is then measured as
\begin{linenomath}\begin{equation}
    \effXY = \frac{\npass}{\npass + \nfail},
\end{equation}\end{linenomath}
where $y$ denotes the reference sample of muon candidates and $x$ is the probe criterion.
The numbers \npass and \nfail correspond to the number of events that pass and fail the test criterion, respectively.

For each of the efficiencies, and in bins of 20\pbinv, fits to the \mll distributions of the passing and failing distributions are performed.
In the fits, the same shapes as described in Section~\ref{sec:effHLT} are used to describe the signal.
In the histograms with passing probes, the background contribution is low and a falling exponential is used.
In the case of failing probes, the nonresonant background is much larger and a more complex analytic function, comprising an exponential at high mass above the \PZ boson resonance and an error function at low mass, is fit.
To ensure a bias-free measurement of \effMGloSta, the outer standalone muon track parameters are used to determine \mll for the passing and failing probes.
Since the resolution of these tracks is much worse, the invariant mass requirement is widened to \mbox{50--130\GeV}.
In the case that, in a given event, the probe muon also fulfills the tag muon requirements,
the tag-and-probe muons are indistinguishable and both muons are used as probes.
Quantitative results for the measurement of the efficiencies are presented in Section~\ref{sec:results}.

\subsection{Acceptance correction}
\label{sec:acceptance}

To determine the true number of \PZ bosons in the visible phase space,
an acceptance correction for losses, or gains, due to the finite resolution of the reconstructed muon tracks is required.
The correction affects the number of reconstructed \PZ bosons itself.
The efficiencies are also affected, primarily in the matching of inner and outer tracks, and, to a lesser extent,
if muon tracks for passing and for failing probes have different resolutions.
The size of the correction is determined from the simulation by comparing the efficiency-corrected number of \PZ bosons as obtained from the measurement with the generated number of \PZ bosons in the visible phase space, as defined for bare leptons after final-state radiation (FSR), but before detector simulation.

For outer muon tracks, resolution effects lead to an acceptance correction of about 1.35\%, which is independent of pileup and constant over the full year of data taking. For inner tracks, the acceptance correction is 0.15\% at low pileup, and it is negligibly small for the \highPU data set. This pileup dependency of 0.15\% is applied as an additional correction, and an uncertainty of 100\% of the correction is assigned.
For a direct cross section measurement, the size of the bias could be further reduced through optimized track selection criteria.
However, for this analysis it suffices that the pileup-independent components of the acceptance correction cancel in the cross section ratio.

\subsection{The L1 trigger corrections}
\label{sec:prefire}

The term ``prefiring'' describes the effect that a trigger decision is assigned to a bunch crossing preceding the one in which the collision actually took place.
In the CMS experiment, the triggering and readout of events in adjacent bunch crossings is vetoed in the trigger logic.
However, due to the limited time resolution of the muon system, the assignment of muon candidates to bunch crossings can be wrong, and thus lead to a loss of good events, \ie, a trigger inefficiency.
Since the tag-and-probe efficiency measurement is insensitive to this effect, the inefficiency due to prefiring is measured in a dedicated analysis.
During the 2017 data taking, measurable prefiring occurred at nonnegligible rates for the L1 muon and ECAL triggers~\cite{CMS:TRG-17-001,CMS:MUO-19-001}.
For the L1 muon trigger, a correction for trigger inefficiency of 0.6\% was found, independent of pileup and time.
In contrast, losses due to prefiring of the ECAL require a pileup-dependent correction of 0.05--0.2\% for the pileup range 0--50.
The prefiring from ECAL triggers is caused mainly by initial or final state radiation, pileup jets, or the underlying event.
The impact on the \lowPU data was somewhat larger due to the lower ECAL trigger thresholds, and for the \lowPU data a correction of 0.6\% is applied.

\section{Results and uncertainties}
\label{sec:results}

The procedures described above to measure the number of reconstructed \PZ bosons and their efficiencies are applied to the data in bins of 20\pbinv.
Since the amount of data at the end of a fill does not usually add up to 20\pbinv, the last bin is included as long as it contains more than 10\pbinv.
In case the last bin contains ${<}10\pbinv$ it is merged with the second to last.
Altogether, in 2017 about 2000 such bins are defined.

\subsection{Normalized \texorpdfstring{\PZ}{Z} boson rate}

\begin{figure}[!t]
\centering
\includegraphics[width=0.48\textwidth]{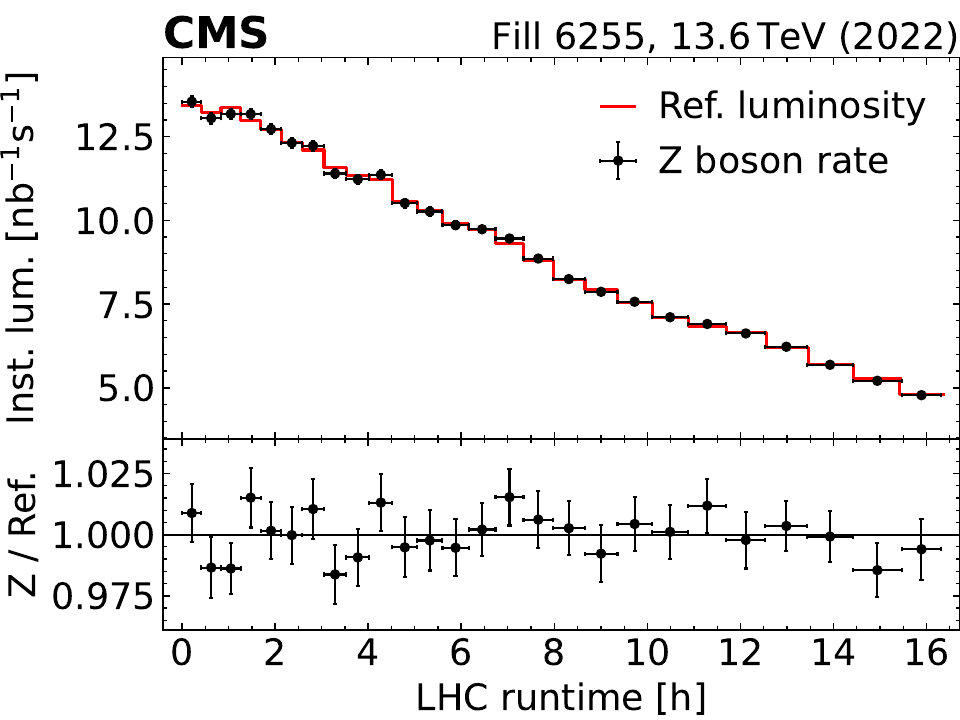}%
\hfill%
\includegraphics[width=0.48\textwidth]{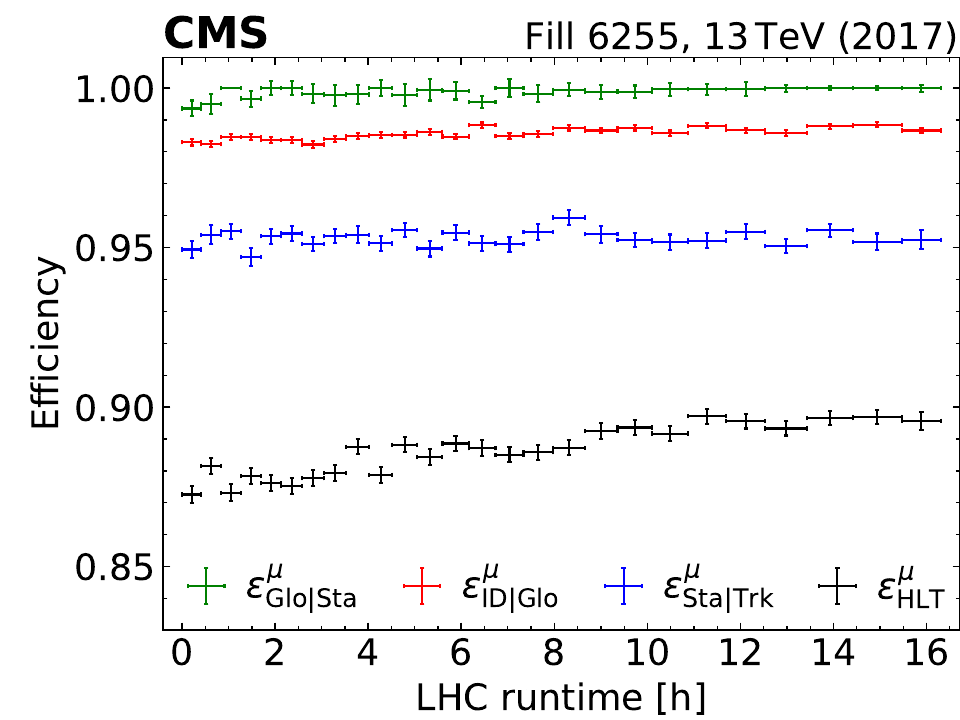}
\caption{%
    \cmsLeft:\ the efficiency-corrected \PZ boson rate, compared to the reference luminosity measurement, in the LHC fill 6255, recorded on September 29, 2017~\cite{CMS:LUM-17-004}.
    Each bin corresponds to about 20\pbinv, as determined by the reference measurement.
    For shape comparison, the integrated \PZ boson rate is normalized to the reference integrated luminosity.
    The panel at the bottom shows the ratio of the two measurements. The vertical error bars show the statistical uncertainty in the \PZ boson rate.
    \cmsRight:\ the measured single-muon efficiencies as functions of time for the same LHC fill.
    The vertical error bars show the statistical uncertainty in the efficiency.
}
\label{fig:rate}
\end{figure}

In Fig.~\ref{fig:rate}, the measured \PZ boson rate and efficiencies are shown for the data recorded during a typical LHC fill of the \highPU data-taking period in 2017.
In this fill, \pp collision data were recorded continuously for about 16 hours.
An integrated luminosity of about 515\pbinv was accumulated, corresponding to 25 bins of 20\pbinv each, and the last bin contains the remaining 15\pbinv of data.
The instantaneous luminosity decreased from initially 15\nbsinv, corresponding to a pileup of about 50, to about one third of the initial value.
In Fig.~\ref{fig:rate} (\cmsleft), a comparison between the conventional measurement of the recorded luminosity and the measurement using the \PZ boson rate is shown.
The integral of the measured \PZ boson rate is normalized to the integral of the reference luminosity.
The shapes of the two independent measurements agree very well.
In Fig.~\ref{fig:rate} (\cmsright), the muon trigger and identification efficiencies, \effMHLT and \effMID, separated into its different components, as applied to the respective time intervals, are presented.
In particular, a significant dependence on time, and thus on pileup, is seen for the HLT muon efficiency for which a rise by about 3\% is measured as the pileup decreases in the course of the fill.

\begin{figure*}[!t]
\centering
\includegraphics[width=\textwidth]{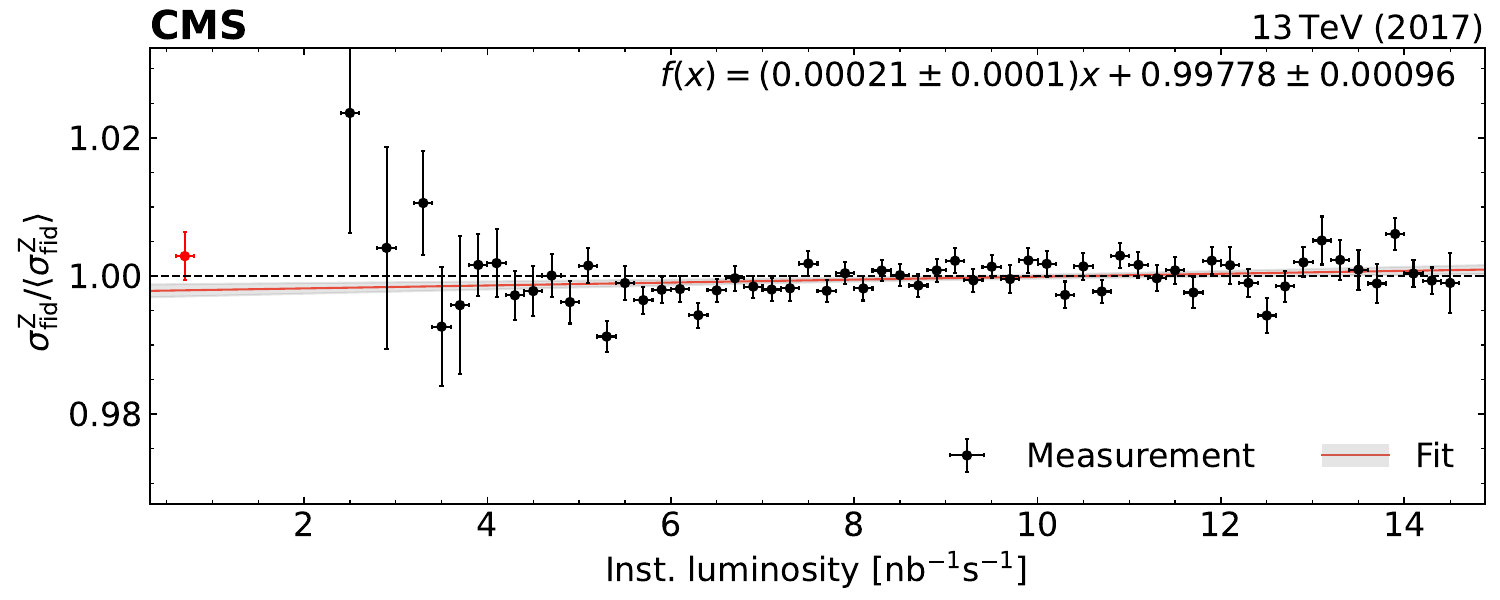}
\caption{%
    Fiducial \PZ boson production cross section as a function of the instantaneous recorded luminosity, normalized to the average measured cross section.
    In each point, multiple measurements of the delivered \PZ boson rates are combined, the error bars correspond to the statistical uncertainties of the \PZ  boson rate measurement.
    The leftmost point, highlighted in red, corresponds to the \lowPU data.
    The result of a fit to a linear function is shown as a red line and the statistical uncertainties are covered by the gray band.
}
\label{fig:linearity}
\end{figure*}

To compare the relative linearity between the measurement of the \PZ boson rate and the CMS reference luminosity, the fiducial cross section for \PZ boson production, normalized to the average \PZ boson cross section, is studied as a function of the instantaneous luminosity.
The result is shown in Fig.~\ref{fig:linearity}, where the average instantaneous luminosity in each 20\pbinv bin is used to assign an instantaneous luminosity bin from which the average cross section is obtained.
The straight-line fit to the data yields a value of 0.2\% below unity for the intercept with the $y$-axis at low pileup. This value gives an estimate of the agreement between \PZ boson counting and reference luminosity measurement in the extrapolation from the low to the high pileup data.

\subsection{Measurement of the absolute luminosity}

Using Eq.~\eqref{eqn:lumiratio}, the integrated luminosity of the \highPU data, referred to in the following as ``\PZ luminosity'', is determined from the integrated luminosity of the \lowPU data, using the ratio of the number of \PZ bosons recorded during the two periods, corrected for reconstruction and trigger efficiencies as determined in intervals of 20\pbinv.
In the ratio, all correlated uncertainties cancel, as detailed in the following section.

\begin{figure}[!t]
\centering
\includegraphics[width=0.48\textwidth]{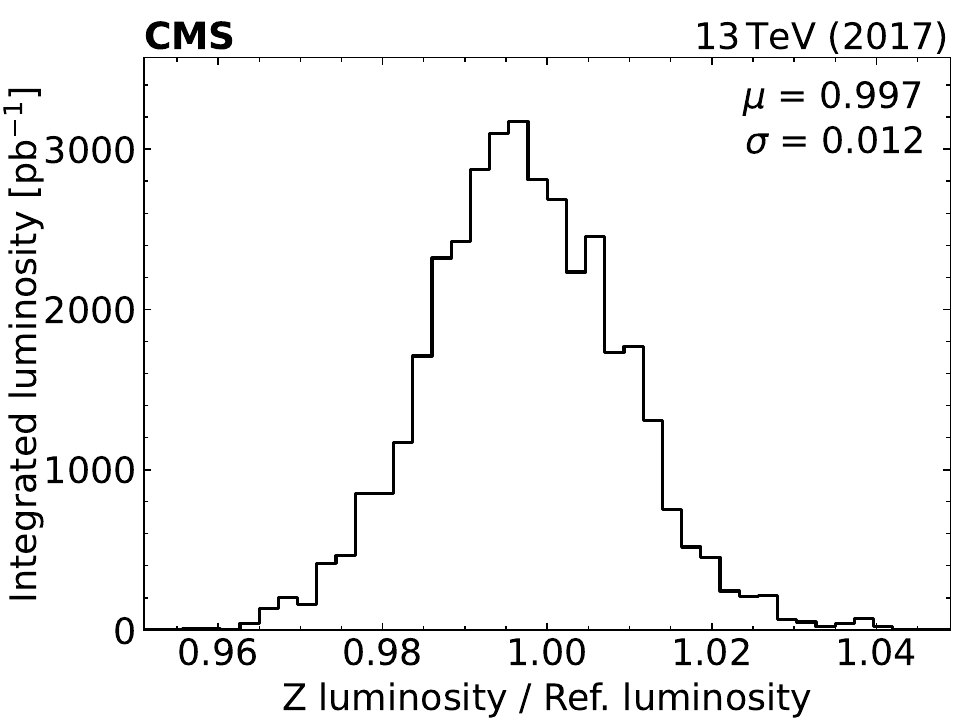}
\caption{%
    Distribution of the ratio of integrated luminosities between \PZ boson counting and the reference luminometer. The entries, each corresponding to one interval of 20\pbinv of \highPU data, are weighted with the respective measured luminosity.
}
\label{fig:zlumiratiohist}
\end{figure}

\begin{figure*}[!t]
\centering
\includegraphics[width=\textwidth]{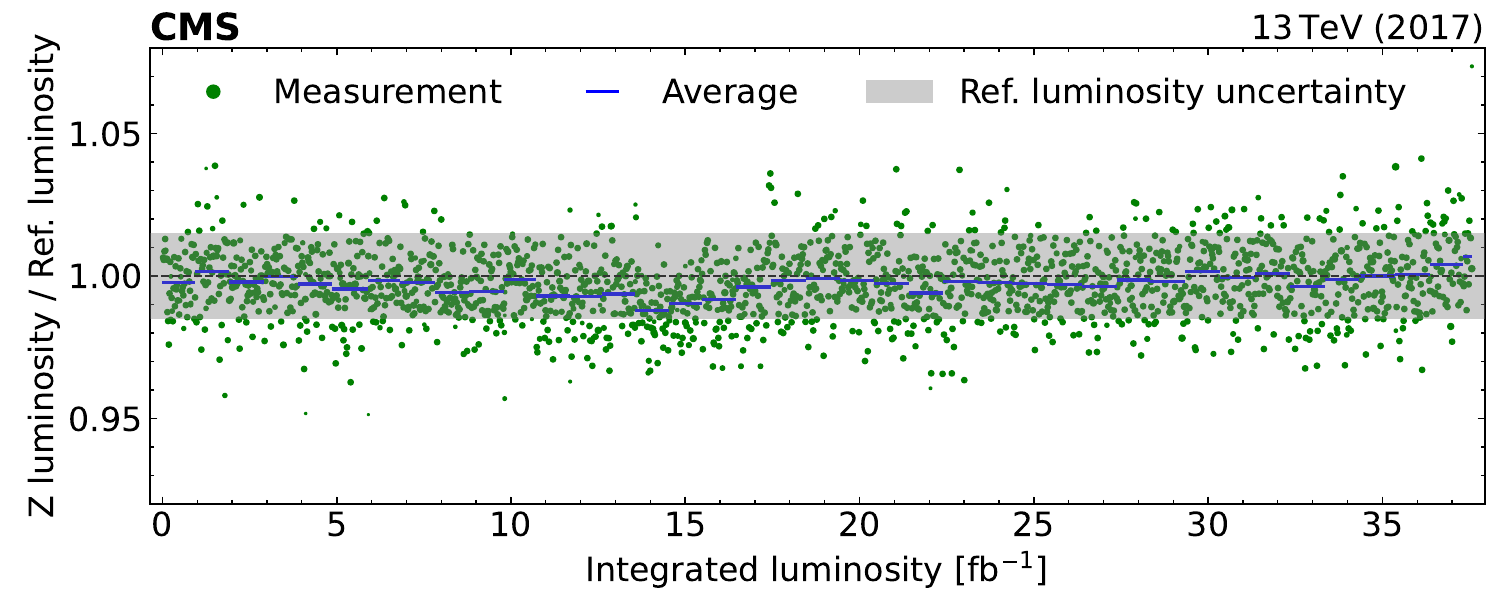}
\caption{%
    The luminosity as measured from \PZ bosons divided by the reference luminosity as a function of the integrated luminosity for the 2017 \highPU data.
    Each green point represents the ratio from one measurement of the number of \PZ bosons.
    The blue lines show the averages of 50 consecutive measurements, corresponding to an average of 1\fbinv of data.
    The gray band has a width of 1.5\%, corresponding to the uncertainty in the ratio of the integrated reference luminosities from the \lowPU to the one of \highPU~\cite{CMS:LUM-17-004}.
}
\label{fig:zlumiratio}
\end{figure*}

In Fig.~\ref{fig:zlumiratiohist}, the distribution of the ratios between the \PZ luminosity and the reference luminosity as obtained from the CMS luminosity systems is shown. Each entry in the histogram corresponds to an interval of 20\pbinv in the \highPU data recorded in 2017.
The central values of both measurements are in good agreement with a difference of 0.3\%.
The standard deviation of about 1.2\% is predominantly of statistical nature,
and close to the expectation for the pure statistical uncertainty of about 12\,000 \PZ boson candidates reconstructed in intervals of 20\pbinv each.
The ratio of \PZ luminosity and reference luminosity as a function of the integrated luminosity is shown in Fig.~\ref{fig:zlumiratio}.
This figure shows a good stability of the \PZ luminosity measurement over the full year. No significant patterns in time are observed.

\subsection{Statistical and systematic uncertainties, and additional cross checks}
\label{sec:systematics}

The uncertainties in the analysis were studied with the focus on the ratio $r=\NZ_\highPU/\NZ_\lowPU$ of the \PZ boson counts between two data samples in 2017 as presented in Eq.~\eqref{eqn:lumiratio}.
The full list of considered sources of uncertainty in the cross sections and their ratio is given in Table~\ref{tab:SystSummary}, and described in the following.

\begin{table*}[!ht]
\centering
\topcaption{%
    Summary of the uncertainties in the number of delivered \PZ bosons in the 2017 \highPU and \lowPU data, and their ratio.
    The symbol $\delta$ denotes the relative uncertainty, \ie, $\delta x = \Delta x / x$.
    The systematic and statistical uncertainties are added in quadrature to obtain the total uncertainty.
}
\label{tab:SystSummary}
\renewcommand{\arraystretch}{1.1}
\begin{tabular}{lcccccc}
    & $\delta\NZ_\highPU$ [\%] & $\delta\NZ_\lowPU$ [\%] & $\delta \Big(\NZ_\highPU/\NZ_\lowPU\Big)$ [\%] \\
    \hline
    HLT correlation \CHLT & $\pm0.1$ & $\pm0.06$ & $\pm0.04$ \\
    Dimuon correlation \CID & $\pm0.00$ & $\mp0.01$ & $\pm0.01$ \\
    Inner-outer track correlation \CIO & $\pm0.01$ & $\mp0.01$ & $\pm0.01$ \\
    Inner track resolution & $\pm0.01$ & $\pm0.16$ & $\mp0.15$ \\
    Outer track resolution & $\pm 1.35$ & $\pm 1.36$ & $\mp0.01$ \\
    L1 muon prefiring & $\pm0.15$ & $\pm0.15$ & $0$ \\
    ECAL prefiring & $\pm0.04$ & $\pm0.14$ & $\mp0.10$ \\
    Signal modeling up & $-0.63$ & $-0.75$ & $+0.19$ \\
    Signal modeling down & $+0.51$ & $+0.71$ & $-0.21$ \\
    Background modeling up & $-0.15$ & $-0.31$ & $+0.16$ \\
    Background modeling down & $-0.09$ & $-0.05$ & $-0.04$ \\[\cmsTabSkip]
    Systematic up & $+1.45$ & $+1.56$ & $+0.31$ \\
    Systematic down & $-1.50$ & $-1.60$ & $-0.28$ \\
    Statistical & $\pm0.03$ & $\pm0.35$ & $\pm0.35$ \\[\cmsTabSkip]
    Total up & $+1.45$ & $+1.60$ & $+0.47$ \\
    Total down & $-1.50$ & $-1.64$ & $-0.45$ \\
\end{tabular}
\end{table*}

Statistical uncertainties are driven by the number of available \PZ bosons and also include the statistical uncertainty in the efficiencies.
As mentioned above, in one interval of 20\pbinv, about 12\,000 \PZ bosons candidates with two muons in the final state are available,
leading to an average statistical uncertainty of 1.17\%.
For all intervals combined, the statistical uncertainty for the full 2017 \highPU data is negligibly small.
The \lowPU data set corresponds to an integrated luminosity of about 200\pbinv, and this contributes a statistical uncertainty of about 0.35\%.

As discussed in Section~\ref{sec:HLTcor}, the correction factor for correlations in the trigger efficiencies of the two muons \CHLT is determined from data and simulation; it is about 0.1\% above unity for the \highPU sample, consistently for data and MC simulation.
The uncertainty in \CHLT is assigned to be 100\% of the correction.

Possible correlations between the two identified mu\-ons and imperfect factorization of muon identification and reconstruction efficiencies were discussed in Section~\ref{sec:effrec}. The simulation shows negligible effects, and corrections at the level of 0.01\% are applied.
The corresponding uncertainties are estimated to be 100\% of the correction.

The limited resolution of the reconstructed muon tracks leads to a bias in the measurement, as described in Section~\ref{sec:acceptance}.
The bias from the inner track resolution is smaller, but pileup dependent, and remains in the ratio with a magnitude of 0.15\%.
The outer track resolution leads to a large bias, but is mostly pileup independent and cancels in the ratio.
A correction is derived from simulation and two independent sources of uncertainty, estimated to be 100\% of the correction each, are assigned each for the inner and outer tracks, respectively.

Systematic uncertainties in the L1 muon prefiring corrections, described in Section~\ref{sec:prefire}, cancel completely in the ratio, whereas the uncertainties due to ECAL prefiring have a different magnitude between the two data sets and cancel only partially.
The remaining uncertainty is estimated to be 20\% of the nominal correction~\cite{CMS:TRG-17-001}.

The extraction of the signal and background contributions was studied using alternative fit models.
For the signal model, the Gaussian resolution function convolved with the histogram template is varied.
First, the histogram template is used alone, \ie, fully relying on the simulation and leaving no further degrees of freedom to the fit.
Secondly, the histogram template is convolved with a Crystal Ball function~\cite{Oreglia:1980cb},
which has four free parameters and gives the fit more freedom to incorporate possible differences between data and simulation.
Thirdly, the histogram template is constructed from generator-level post-FSR leptons, instead of mirroring the selection at detector level.
This template is then convolved with a Crystal Ball function.
The three variations lead to changes in the extracted numbers of \PZ bosons, and the efficiencies, in both the \highPU and \lowPU data sets.
While the trends are correlated, the relative magnitudes are different, and this leads to a significant residual uncertainty.
The envelope of the three variations is taken to quantify this uncertainty.

The two types of background models are varied independently.
For the categories with major background contributions, the Das function~\cite{Das:2016stf}, a wide Gaussian distribution with exponential tails is used as an alternative function, which has four free parameters, as opposed to three for the nominal model.
In the other cases, the falling exponential is substituted by a uniform distribution.

{\tolerance=800
The total systematic uncertainty is obtained by adding the systematic uncertainties listed in Table~\ref{tab:SystSummary} in quadrature.
In combination with the statistical uncertainty of 0.35\%, the total uncertainty to transfer the luminosity from the \lowPU data to the \highPU data in 2017 is
\begin{linenomath}\begin{equation}
    \delta r = {}^{+0.31\%}_{-0.28\%}\syst \pm 0.35\%\stat = {}^{+0.47\%}_{-0.45\%}\,,
\end{equation}\end{linenomath}
where the statistical uncertainty is due to the limited size of the \lowPU data set.
The systematic uncertainty is driven by the uncertainties in the signal modeling, followed by the background modeling and acceptance corrections.
Overall, a total uncertainty of about 0.5\% is obtained.
\par}

\begin{table*}[!t]
\centering
\topcaption{
    Summary of cross checks performed by varying the length of the luminosity interval, the bin width of the \mll histograms, and the range of the fit.
    As in Table~\ref{tab:SystSummary}, the resulting variations of the number of \PZ bosons in the 2017 \highPU and \lowPU data, and their ratio, are shown.
    The $\delta$ denotes the relative variations, \ie, $\delta x = \Delta x / x$.
}
\label{tab:crosschecks}
\renewcommand{\arraystretch}{1.2}
\begin{tabular}{lcccccc}
    & $\delta\NZ_\highPU$ [\%] & $\delta\NZ_\lowPU$ [\%] & $\delta \Big(\NZ_\highPU/\NZ_\lowPU\Big)$ [\%] \\
    \hline
    Lum. bin size 30\pbinv & $-0.05$ & $-0.01$ & $-0.04$ \\
    Lum. bin size 15\pbinv & $+0.04$ & $+0.07$ & $-0.03$ \\
    Mass bin width 1\GeV & $-0.02$ & $-0.06$ & $+0.03$ \\
    Mass bin width 0.25\GeV & $-0.01$ & $+0.01$ & $-0.02$ \\
    Mass range $[50,130]\GeV$ & $+1.25$ & $+1.24$ & $+0.00$ \\
    Mass range $[70,110]\GeV$ & $-2.32$ & $-2.26$ & $-0.05$ \\
\end{tabular}
\end{table*}

Multiple cross-checks were performed to test the robustness of the result.
The size of the luminosity bin was varied from 20 down to 15 and up to 30\pbinv, and negligible differences with respect to the nominal measurement were found.
It was further verified that the measurement is independent of the choice of the bin width chosen for the \mll distribution, by varying it by factors of $1/2$ and 2.
Independence of the result on the chosen fit interval was tested using two alternative ranges:\ a more narrow interval from $\mll \in [70,110]\GeV$ and a wider interval from $\mll \in [50,130]\GeV$.
Both variations have a strong impact on \NZ since the phase space of the measurement changes, but, as expected, the effect cancels almost completely in the ratio.
The results of these cross checks are summarized in Table~\ref{tab:crosschecks}.

\section{Discussion and outlook}
\label{sec:outlook}

With an uncertainty in the transfer factor of about 0.5\% for the 2017 data, this analysis shows that \PZ boson counting can provide an independent and competitive method to extrapolate and integrate luminosity calibrations.
The results from \PZ boson counting are independent of the conventional luminosity measurements.
They can be treated as uncorrelated in combinations, which can lead to significant improvements in the combined uncertainty.

Taking the current precision of 1.7\% for the integrated luminosity in the \lowPU data~\cite{CMS:LUM-17-004}, the integrated luminosity in the \highPU 2017 data could potentially be determined to a precision of better than 1.8\%,
in contrast to the preliminary uncertainty of the reference luminosity measurement of 2.3\%~\cite{CMS:LUM-17-004}.

A unique aspect of \PZ boson counting is that the relevant efficiency corrections as a function of time can be calibrated from the same event sample.
This feature makes the method robust not only against small changes in detector response, but also across different detector configurations.
In general, once a precision measurement of the integrated luminosity is available, such as that for the \lowPU data in 2017, the integrated luminosity for all data recorded at the same center-of-mass energy can be determined using the \PZ boson counting.
However, each transfer between data sets requires detailed studies of the correlations of the muon trigger and the reconstruction efficiencies.

In this paper, the full analysis was presented for the data from 2017, when a dedicated and sufficiently large sample of \lowPU data was recorded.
Under such conditions, a large fraction of the systematic uncertainties cancels in the ratio.
For the most precise CMS measurement of the luminosity to date~\cite{CMS:LUM-17-003}, published for 2016, an extrapolation and integration uncertainty of 0.7\% was reported.
For 2016, no \lowPU data set was recorded.
Further studies on the impact of different detector conditions would be required to extrapolate from the 2016 data set.
If, hypothetically, an extrapolation uncertainty of 0.5\% for \PZ boson counting were achievable also in the 2016 data, the uncertainty of 1.2\% in the total integrated luminosity for 2016 could be improved to 1.1\%.

{\tolerance=800
The dominant contribution to the uncertainty comes from the statistical uncertainty, which is driven by the size of the \lowPU data sample.
The \lowPU data recorded in 2017 correspond to an integrated luminosity of about 200\pbinv.
A significant increase of the sample size, \eg, by a factor 3 or 4, would make the statistical uncertainty negligible.
\par}

In the coming years, during the ongoing LHC \Run3 and beyond, additional measurements and studies on the main systematic uncertainties will be performed,
and that is expected to improve the precision of the method further. Furthermore, the method is expected to contribute substantially to the combination of integrated luminosity measurements for different data sets. In the longer term, pileup conditions of up to 200 \pp collisions per bunch crossing are expected at the HL-LHC~\cite{Dainese:2703572}. In both \Run3 and at the HL-LHC, the uncertainties due to extrapolation from vdM conditions to standard data taking are expected to remain substantial. In such conditions, the method of \PZ boson counting has the potential to provide significant improvements.

\section{Summary}
\label{sec:summary}

The precision measurement of the \PZ boson production rate provides a complementary method to transfer integrated luminosity measurements between data sets.
This study makes use of events with \PZ bosons decaying into a pair of muons.
The data were recorded with the CMS experiment at the CERN LHC in 2017, at a proton-proton center-of-mass energy of 13\TeV.
The integrated luminosity of a larger data sample recorded in 2017 is obtained from that of a smaller data set recorded at lower pileup using the ratio of the efficiency-corrected numbers of \PZ bosons counted in the two data sets.
The full set of efficiencies and correlation correction factors for triggering, reconstruction, and selection are determined in intervals of 20\pbinv from the same \PZ boson data samples.
Monte Carlo simulations are used only to describe the shape of the resonant \PZ boson signal and for the study of possible biases of the method.
A detailed quantitative study of the systematic uncertainties and their dependencies on pileup is performed for the first time.
In the integrated luminosity ratio, the systematic uncertainties cancel almost completely, with the exception of the pileup-dependent effects.
The resulting uncertainty in the ratio is 0.5\%.
With its high precision, the \PZ boson counting is competitive with and independent of conventional methods for the extrapolation and integration of luminosity.

\begin{acknowledgments}
We congratulate our colleagues in the CERN accelerator departments for the excellent performance of the LHC and thank the technical and administrative staffs at CERN and at other CMS institutes for their contributions to the success of the CMS effort. In addition, we gratefully acknowledge the computing centers and personnel of the Worldwide LHC Computing Grid and other centers for delivering so effectively the computing infrastructure essential to our analyses. Finally, we acknowledge the enduring support for the construction and operation of the LHC, the CMS detector, and the supporting computing infrastructure provided by the following funding agencies: SC (Armenia), BMBWF and FWF (Austria); FNRS and FWO (Belgium); CNPq, CAPES, FAPERJ, FAPERGS, and FAPESP (Brazil); MES and BNSF (Bulgaria); CERN; CAS, MoST, and NSFC (China); MINCIENCIAS (Colombia); MSES and CSF (Croatia); RIF (Cyprus); SENESCYT (Ecuador); MoER, ERC PUT and ERDF (Estonia); Academy of Finland, MEC, and HIP (Finland); CEA and CNRS/IN2P3 (France); BMBF, DFG, and HGF (Germany); GSRI (Greece); NKFIH (Hungary); DAE and DST (India); IPM (Iran); SFI (Ireland); INFN (Italy); MSIP and NRF (Republic of Korea); MES (Latvia); LAS (Lithuania); MOE and UM (Malaysia); BUAP, CINVESTAV, CONACYT, LNS, SEP, and UASLP-FAI (Mexico); MOS (Montenegro); MBIE (New Zealand); PAEC (Pakistan); MES and NSC (Poland); FCT (Portugal); MESTD (Serbia); MCIN/AEI and PCTI (Spain); MOSTR (Sri Lanka); Swiss Funding Agencies (Switzerland); MST (Taipei); MHESI and NSTDA (Thailand); TUBITAK and TENMAK (Turkey); NASU (Ukraine); STFC (United Kingdom); DOE and NSF (USA).

\hyphenation{Rachada-pisek} Individuals have received support from the Marie-Curie program and the European Research Council and Horizon 2020 Grant, contract Nos.\ 675440, 724704, 752730, 758316, 765710, 824093, and COST Action CA16108 (European Union); the Leventis Foundation; the Alfred P.\ Sloan Foundation; the Alexander von Humboldt Foundation; the Science Committee, project no. 22rl-037 (Armenia); the Belgian Federal Science Policy Office; the Fonds pour la Formation \`a la Recherche dans l'Industrie et dans l'Agriculture (FRIA-Belgium); the Agentschap voor Innovatie door Wetenschap en Technologie (IWT-Belgium); the F.R.S.-FNRS and FWO (Belgium) under the ``Excellence of Science -- EOS" -- be.h project n.\ 30820817; the Beijing Municipal Science \& Technology Commission, No. Z191100007219010 and Fundamental Research Funds for the Central Universities (China); the Ministry of Education, Youth and Sports (MEYS) of the Czech Republic; the Shota Rustaveli National Science Foundation, grant FR-22-985 (Georgia); the Deutsche Forschungsgemeinschaft (DFG), under Germany's Excellence Strategy -- EXC 2121 ``Quantum Universe" -- 390833306, and under project number 400140256 - GRK2497; the Hellenic Foundation for Research and Innovation (HFRI), Project Number 2288 (Greece); the Hungarian Academy of Sciences, the New National Excellence Program - \'UNKP, the NKFIH research grants K 124845, K 124850, K 128713, K 128786, K 129058, K 131991, K 133046, K 138136, K 143460, K 143477, 2020-2.2.1-ED-2021-00181, and TKP2021-NKTA-64 (Hungary); the Council of Science and Industrial Research, India; the Latvian Council of Science; the Ministry of Education and Science, project no. 2022/WK/14, and the National Science Center, contracts Opus 2021/41/B/ST2/01369 and 2021/43/B/ST2/01552 (Poland); the Funda\c{c}\~ao para a Ci\^encia e a Tecnologia, grant CEECIND/01334/2018 (Portugal); the National Priorities Research Program by Qatar National Research Fund; MCIN/AEI/10.13039/501100011033, ERDF ``a way of making Europe", and the Programa Estatal de Fomento de la Investigaci{\'o}n Cient{\'i}fica y T{\'e}cnica de Excelencia Mar\'{\i}a de Maeztu, grant MDM-2017-0765 and Programa Severo Ochoa del Principado de Asturias (Spain); the Chulalongkorn Academic into Its 2nd Century Project Advancement Project, and the National Science, Research and Innovation Fund via the Program Management Unit for Human Resources \& Institutional Development, Research and Innovation, grant B05F650021 (Thailand); the Kavli Foundation; the Nvidia Corporation; the SuperMicro Corporation; the Welch Foundation, contract C-1845; and the Weston Havens Foundation (USA).
\end{acknowledgments}

\bibliography{auto_generated}
\cleardoublepage \appendix\section{The CMS Collaboration \label{app:collab}}\begin{sloppypar}\hyphenpenalty=5000\widowpenalty=500\clubpenalty=5000
\cmsinstitute{Yerevan Physics Institute, Yerevan, Armenia}
{\tolerance=6000
A.~Hayrapetyan, A.~Tumasyan\cmsAuthorMark{1}\cmsorcid{0009-0000-0684-6742}
\par}
\cmsinstitute{Institut f\"{u}r Hochenergiephysik, Vienna, Austria}
{\tolerance=6000
W.~Adam\cmsorcid{0000-0001-9099-4341}, J.W.~Andrejkovic, T.~Bergauer\cmsorcid{0000-0002-5786-0293}, S.~Chatterjee\cmsorcid{0000-0003-2660-0349}, K.~Damanakis\cmsorcid{0000-0001-5389-2872}, M.~Dragicevic\cmsorcid{0000-0003-1967-6783}, A.~Escalante~Del~Valle\cmsorcid{0000-0002-9702-6359}, P.S.~Hussain\cmsorcid{0000-0002-4825-5278}, M.~Jeitler\cmsAuthorMark{2}\cmsorcid{0000-0002-5141-9560}, N.~Krammer\cmsorcid{0000-0002-0548-0985}, D.~Liko\cmsorcid{0000-0002-3380-473X}, I.~Mikulec\cmsorcid{0000-0003-0385-2746}, J.~Schieck\cmsAuthorMark{2}\cmsorcid{0000-0002-1058-8093}, R.~Sch\"{o}fbeck\cmsorcid{0000-0002-2332-8784}, D.~Schwarz\cmsorcid{0000-0002-3821-7331}, M.~Sonawane\cmsorcid{0000-0003-0510-7010}, S.~Templ\cmsorcid{0000-0003-3137-5692}, W.~Waltenberger\cmsorcid{0000-0002-6215-7228}, C.-E.~Wulz\cmsAuthorMark{2}\cmsorcid{0000-0001-9226-5812}
% ...
\par}
\vskip\cmsinstskip
\dag:~Deceased\\
$^{1}$Also at Yerevan State University, Yerevan, Armenia\\
$^{2}$Also at TU Wien, Vienna, Austria\\
% ...
\end{sloppypar}
%%% END EDITABLE REGION %%%
% skeleton_end
\end{document}